\documentclass[preprintnumbers,amsmath,amssymbm,prd]{revtex4}
\usepackage{epsfig}
\usepackage{graphicx}

\begin{document}
\title{Numerical evidence for universality in the excited instability spectrum of magnetically
charged Reissner-Nordstr\"om black holes}
\author{Shahar Hod}
\address{The Ruppin Academic Center, Emeq Hefer 40250, Israel}
\address{ }
\address{The Hadassah Institute, Jerusalem 91010, Israel}
\date{\today}

\begin{abstract}
\ \ \ It is well-known that the SU(2) Reissner-Nordstr\"om
black-hole solutions of the Einstein-Yang-Mills theory are
characterized by an infinite set of unstable (imaginary) eigenvalues
$\{\omega_n(T_{\text{BH}})\}_{n=0}^{n=\infty}$ (here $T_{\text{BH}}$
is the black-hole temperature). In this paper we analyze the excited
instability spectrum of these magnetically charged black holes. The
numerical results suggest the existence of a universal behavior for
these black-hole excited eigenvalues. In particular, we show that
unstable eigenvalues in the regime $\omega_n\ll T_{\text{BH}}$ are
characterized, to a very good degree of accuracy, by the simple
universal relation $\omega_n(r_+-r_-)={\text{constant}}$, where
$r_{\pm}$ are the horizon radii of the black hole.
\end{abstract}
\bigskip
\maketitle


\section{Introduction}

The familiar U(1) Reissner-Nordstr\"om spacetime is known to
describe a {\it stable} black-hole solution of the coupled
Einstein-Maxwell equations \cite{Mon} and the coupled
Einstein-Maxwell-scalar equations \cite{Hods}. Yasskin \cite{Yas}
has proved that the Einstein-Yang-Mills theory also admits an
explicit black-hole solution which is described by the magnetically
charged SU(2) Reissner-Nordstr\"om spacetime. However, the SU(2)
Reissner-Nordstr\"om black-hole solution of the coupled
Einstein-Yang-Mills equations is known to be {\it unstable}
\cite{Str,BizWal,Hodplb1}. In fact, it was proved in \cite{Mas} that
the magnetically charged Reissner-Nordstr\"om black-hole spacetime
is characterized by an {\it infinite} family of unstable (growing in
time) perturbation modes.

The recent numerical work of Rinne \cite{Oliv} has revealed that
these unstable SU(2) Reissner-Nordstr\"om black-hole spacetimes play
the role of approximate \cite{Noteaprr} codimension-two intermediate
attractors (that is, nonlinear critical solutions \cite{GunMar}) in
the dynamical gravitational collapse of the Yang-Mills field
\cite{Noteot}. In particular, this interesting numerical study
\cite{Oliv} has explicitly demonstrated that, during a near-critical
evolution of the Yang-Mills field, the time spent in the vicinity of
an unstable SU(2) Reissner-Nordstr\"om black-hole solution is
characterized by the critical scaling law \cite{Notepp}
\begin{equation}\label{Eq1}
\tau=\text{const}-\gamma\ln|p-p^*|\  .
\end{equation}
Interestingly, the critical exponents of the scaling law (\ref{Eq1})
are directly related to the characteristic instability eigenvalues
of the corresponding SU(2) Reissner-Nordstr\"om black holes
\cite{Oliv}:
\begin{equation}\label{Eq2}
\gamma=1/\omega_{\text{instability}}\  .
\end{equation}
It is therefore of physical interest to explore the instability
spectrum $\{\omega_n\}_{n=0}^{n=\infty}$ of the SU(2)
Reissner-Nordstr\"om black holes. Indeed, Rinne \cite{Oliv} has
recently computed numerically the characteristic unstable
eigenvalues of these magnetically charged black-hole solutions of
the Einstein-Yang-Mills theory \cite{NoteRin}.

In the present paper we shall analyze these numerically computed
black-hole eigenvalues in an attempt to identify a possible hidden
pattern which characterizes the black-hole instability spectrum. As
we shall show below, the numerical results indeed suggest the
existence of a universal behavior for these black-hole unstable
eigenvalues.

\section{Description of the system}

The Reissner-Nordstr\"om black-hole solution of the
Einstein-Yang-Mills theory with unit magnetic charge is described by
the line element \cite{Yas}
\begin{equation}\label{Eq3}
ds^2=-\Big(1-{{2m}\over{r}}\Big)dt^2+\Big(1-{{2m}\over{r}}\Big)^{-1}dr^2+r^2(d\theta^2+\sin^2\theta
d\phi^2)\ ,
\end{equation}
where the mass function $m=m(r)$ is given by \cite{Noteunit}
\begin{equation}\label{Eq4}
m(r)=M-{{1}\over{2r}}\  .
\end{equation}
The black-hole temperature is given by
\begin{equation}\label{Eq5}
T_{\text{BH}}={{r_+-r_-}\over{4\pi r^2_+}}\  ,
\end{equation}
where
\begin{equation}\label{Eq6}
r_{\pm}=M\pm\sqrt{M^2-1}\
\end{equation}
are the (outer and inner) horizons of the black hole.

Linearized perturbations $\xi(r) e^{-i\omega t}$ \cite{Noteuns} of
the magnetically charged black-hole spacetime are governed by the
Schr\"odinger-like wave equation \cite{Bizw}
\begin{equation}\label{Eq7}
\Big\{{{d^2}\over{dx^2}}+\omega^2+{1 \over {r^2}}\Big[1-{{2m(r)}
\over r}\Big] \Big\}\xi=0\  ,
\end{equation}
where the ``tortoise" radial coordinate $x$ is defined by the
relation \cite{Notehor}
\begin{equation}\label{Eq8}
dx/dr=[1-2m(r)/r]^{-1}\  .
\end{equation}
Well-behaved (spatially bounded) perturbation modes are
characterized by the boundary conditions
\begin{equation}\label{Eq9}
\xi(x\to -\infty)\sim e^{|\omega|x}\to 0\
\end{equation}
and
\begin{equation}\label{Eq10}
\xi(x\to\infty)\sim xe^{-|\omega|x}\to 0\  ,
\end{equation}
where $\omega=i|\omega|$. As shown in \cite{Mas,BizWal}, these
boundary conditions single out a discrete set of unstable
($\Im\omega>0$) black-hole eigenvalues
$\{\omega_n(r_+)\}_{n=0}^{n=\infty}$.

\section{Numerical evidence for universality in the excited instability spectrum}

Most recently, Rinne \cite{Oliv} computed numerically the first
three instability eigenvalues which characterize the SU(2)
Reissner-Nordstr\"om black-hole solutions of the coupled
Einstein-Yang-Mills equations. We have examined these numerically
computed eigenvalues in an attempt to reveal a possible hidden
pattern which characterizes the black-hole instability spectrum.

In Table \ref{Table1} we present the first excited instability
eigenvalues $\{\omega_1(r_+)\}$ of the magnetically charged SU(2)
Reissner-Nordstr\"om black holes. In particular, we display the
dimensionless ratio $\omega_1(r_+)/\pi T_{\text{BH}}$, where the
black-hole temperature $T_{\text{BH}}$ is given by (\ref{Eq5}). We
also display the ratio between the dimensionless quantity
$\omega_1(r_+)\times(r_+-r_-)$ for generic SU(2)
Reissner-Nordstr\"om black holes and the corresponding quantity
$\omega_1(r_+=10)\times(10-1/10)$ for the weakly-magnetized
Reissner-Nordstr\"om black hole with $r_+=10$ \cite{Notelarg}.
Remarkably, the numerical data presented in Table \ref{Table1}
reveals that the black-hole instability eigenvalues in the regime
$\omega_1(r_+)/T_{\text{BH}}\ll1$ are characterized, to a good
degree of accuracy, by the universal relation \cite{Notelam1}
\begin{equation}\label{Eq11}
\omega_1(r_+-r_-)=\lambda_1\ \ \ ; \ \ \
\lambda_1={\text{constant}}\ .
\end{equation}

In order to support this intriguing finding, we display in Table
\ref{Table2} the second excited instability eigenvalues
$\{\omega_2(r_+)\}$ of the SU(2) Reissner-Nordstr\"om black holes.
Remarkably, the numerical data presented in Table \ref{Table2}
provide compelling evidence for the validity of the suggested
universal behavior of the black-hole instability eigenvalues in the
regime $\omega_2(r_+)/T_{\text{BH}}\ll1$. In particular, one finds
\cite{Notelam2}
\begin{equation}\label{Eq12}
\omega_2(r_+-r_-)=\lambda_2\ \ \ ; \ \ \
\lambda_2={\text{constant}}\ .
\end{equation}

\begin{table}[htbp]
\centering
\begin{tabular}{|c|c|c|}
\hline $r_+$ & \ $\omega_1(r_+)/\pi T_{\text{BH}}$\ \ \ & \ \
${{\omega_1(r_+)\times(r_+-r_-)}\over{\omega_1(r_+=10)\times(10-1/10)}}$\ \ \ \\
\hline
\ \ 9.0\ \ \ &\ \ $9.86\times 10^{-2}$\ \ \ &\ \ 0.999\\
\ \ 8.0\ \ \ &\ \ $9.91\times 10^{-2}$\ \ \ &\ \ 0.999\\
\ \ 7.0\ \ \ &\ \ $9.99\times 10^{-2}$\ \ \ &\ \ 0.998\\
\ \ 6.0\ \ \ &\ \ $1.01\times 10^{-1}$\ \ \ &\ \ 0.996\\
\ \ 5.0\ \ \ &\ \ $1.04\times 10^{-1}$\ \ \ &\ \ 0.993\\
\ \ 4.0\ \ \ &\ \ $1.08\times 10^{-1}$\ \ \ &\ \ 0.987\\
\ \ 3.0\ \ \ &\ \ $1.18\times 10^{-1}$\ \ \ &\ \ 0.973\\
\ \ 2.0\ \ \ &\ \ $1.58\times 10^{-1}$\ \ \ &\ \ 0.925\\
\ \ 1.5\ \ \ &\ \ $2.57\times 10^{-1}$\ \ \ &\ \ 0.824\\
\ \ 1.2\ \ \ &\ \ $6.04\times 10^{-1}$\ \ \ &\ \ 0.586\\
\hline
\end{tabular}
\caption{The instability eigenvalues of SU(2) Reissner-Nordstr\"om
black holes. The data shown refers to the first excited eigenvalues
$\{\omega_1(r_+)\}$ of these magnetically charged black holes. We
display the dimensionless ratio $\omega_1(r_+)/\pi T_{\text{BH}}$,
where $T_{\text{BH}}$ is the black-hole temperature. Also shown is
the ratio between the dimensionless quantity
$\omega_1(r_+)\times(r_+-r_-)$ for generic SU(2)
Reissner-Nordstr\"om black holes and the corresponding quantity
$\omega_1(r_+=10)\times(10-1/10)$ for the weakly-magnetized
Reissner-Nordstr\"om black hole with $r_+=10$ \cite{Notelarg}. One
finds that the instability eigenvalues in the regime
$\omega_1(r_+)/\pi T_{\text{BH}}\lesssim 0.1$ are characterized, to
a good degree of accuracy, by the universal relation
$\omega_1(r_+-r_-)={\text{constant}}$.} \label{Table1}
\end{table}

\begin{table}[htbp]
\centering
\begin{tabular}{|c|c|c|}
\hline $r_+$ & \ $\omega_2(r_+)/\pi T_{\text{BH}}$\ \ \ & \ \
${{\omega_2(r_+)\times(r_+-r_-)}\over{\omega_2(r_+=10)\times(10-1/10)}}$\ \ \ \\
\hline
\ \ 9.0\ \ \ &\ \ $2.90\times 10^{-3}$\ \ \ &\ \ 1.011\\
\ \ 8.0\ \ \ &\ \ $2.95\times 10^{-3}$\ \ \ &\ \ 1.021\\
\ \ 7.0\ \ \ &\ \ $3.00\times 10^{-3}$\ \ \ &\ \ 1.029\\
\ \ 6.0\ \ \ &\ \ $3.06\times 10^{-3}$\ \ \ &\ \ 1.034\\
\ \ 5.0\ \ \ &\ \ $3.15\times 10^{-3}$\ \ \ &\ \ 1.038\\
\ \ 4.0\ \ \ &\ \ $3.31\times 10^{-3}$\ \ \ &\ \ 1.040\\
\ \ 3.0\ \ \ &\ \ $3.68\times 10^{-3}$\ \ \ &\ \ 1.041\\
\ \ 2.0\ \ \ &\ \ $5.17\times 10^{-3}$\ \ \ &\ \ 1.039\\
\ \ 1.5\ \ \ &\ \ $9.37\times 10^{-3}$\ \ \ &\ \ 1.034\\
\ \ 1.2\ \ \ &\ \ $3.03\times 10^{-2}$\ \ \ &\ \ 1.011\\
\hline
\end{tabular}
\caption{The instability eigenvalues of SU(2) Reissner-Nordstr\"om
black holes. The data shown refers to the second excited eigenvalues
$\{\omega_2(r_+)\}$ of these magnetically charged black holes. We
display the dimensionless ratio $\omega_2(r_+)/\pi T_{\text{BH}}$,
where $T_{\text{BH}}$ is the black-hole temperature. Also shown is
the ratio between the dimensionless quantity
$\omega_2(r_+)\times(r_+-r_-)$ for generic SU(2)
Reissner-Nordstr\"om black holes and the corresponding quantity
$\omega_2(r_+=10)\times(10-1/10)$ for the weakly-magnetized
Reissner-Nordstr\"om black hole with $r_+=10$ \cite{Notelarg}. One
finds that the instability eigenvalues in the regime
$\omega_2(r_+)/T_{\text{BH}}\ll1$ are characterized, to a good
degree of accuracy, by the universal relation
$\omega_2(r_+-r_-)={\text{constant}}$.} \label{Table2}
\end{table}

\newpage

\section{Summary}

The U(1) Reissner-Nordstr\"om black holes are known to be stable
within the framework of the coupled Einstein-Maxwell theory
\cite{Mon,Hods}. This stability property of the black holes
manifests itself in the form of an infinite spectrum of damped
quasi-normal resonances \cite{Leav}. To the best of our knowledge,
for generic U(1) Reissner-Nordstr\"om black holes, there is no
simple universal formula which describes the infinite family of
these damped black-hole quasi-normal resonances.

On the other hand, the SU(2) Reissner-Nordstr\"om black holes are
known to be unstable within the framework of the coupled
Einstein-Yang-Mills theory \cite{Str,BizWal,Hodplb1}. This
instability property of the magnetically charged black holes
manifests itself in the form of an infinite spectrum of
exponentially growing black-hole resonances \cite{Mas}. In this
paper we have provided compelling numerical evidence that the
infinite family of these unstable black-hole resonances can be
described, to a very good degree of accuracy, by the simple {\it
universal} formula
\begin{equation}\label{Eq13}
\omega_n(r_+-r_-)={\text{constant}}_n\ \ \ \text{for} \ \ \
\omega_n\ll T_{\text{BH}}\  .
\end{equation}
We believe that it would be highly interesting to find an analytical
explanation for this numerically suggested universal behavior.

\bigskip
\noindent
{\bf ACKNOWLEDGMENTS}
\bigskip

This research is supported by the Carmel Science Foundation. I would
like to thank Oliver Rinne for sharing with me his numerical data. I
would also like to thank Yael Oren, Arbel M. Ongo, Ayelet B. Lata,
and Alona B. Tea for helpful discussions.


\begin{thebibliography}{99}
\bibitem{Notepp} Here $|p-p^*|$ is a measure for the distance of the initial data from
the threshold (critical) solution \cite{GunMar}.
\bibitem{Noteunit} We use natural units in which $G=c=\hbar=1$.
\bibitem{Noteuns} Note that unstable (growing in time) modes are characterized by
$\Im\omega>0$.
\bibitem{Notehor} Note that the near-horizon limit $r \to r_+$ corresponds to
$x \to -\infty$, whereas the large-$r$ limit $r\to\infty$
corresponds to $x\to\infty$.
\bibitem{Notelarg} The weakly-magnetized SU(2) Reissner-Nordstr\"om black hole
with horizon radius $r_+=10$ is the largest black-hole solution
studied numerically in \cite{Oliv}.
\end{thebibliography}
\end{document}